\documentclass[aps]{revtex4}
\usepackage{amssymb}
\usepackage{makeidx}
    \makeindex
\usepackage{times}
\usepackage{tabularx}
\usepackage{booktabs}
\usepackage{epsfig}

\begin{document}

\title{Lectures on Quantum Information\\Chapter 1: The separability versus entanglement problem}
\author{Aditi Sen(De)\(^1\), Ujjwal Sen\(^1\), Maciej Lewenstein\(^{1, \ast}\), and Anna Sanpera\(^{2, \ast}\)}
\affiliation{\(^1\)ICFO-Institut de Ci\`encies Fot\`oniques, E-08034 Barcelona, Spain\\
\(^2\)Grup de F\'isica Te\`orica, Universitat Aut\`onoma de Barcelona, E-08193 Bellaterra, Spain}

\maketitle



\tableofcontents





\section{Introduction}
Quantum theory, formalised in the first few decades of the 20\(^{\mbox{th}}\)century, contains elements that are radically different 
from the classical description of Nature. 
An important aspect in these fundamental differences
is the existence of quantum correlations in the quantum formalism. 
In the classical description of Nature, if a system is formed by  
different subsystems, 
complete knowledge of the whole system
implies that 
the sum of the information of the subsystems makes up the complete information for the whole
system. 
This is no longer true in the quantum formalism. In the quantum world, 
there exists states of composite systems 
for which we might have the complete information, while our knowledge about
the subsystems might be completely random. 
One may reach some paradoxical conclusions if one applies a classical
description to states which have characteristic quantum signatures.

During the last decade, it was realized that these 
fundamentally nonclassical  states, also denoted as ``entangled states'',
can provide us with something else than just paradoxes. 
They may be \emph{used} to perform tasks that cannot be acheived
with classical states.  
As benchmarks of this turning point in our view of such nonclassical states, 
one might mention  the spectacular discoveries of 
(entanglement-based) quantum cryptography (1991) \cite{Ekert},  quantum dense coding (1992) \cite{BW}, and 
quantum teleportation (1993)
\cite{BBCJPW}. 


In this chapter, we will focus on bipartite composite systems. We will 
define formally what entangled states are, present 
some important criteria to discriminate entangled states from separable ones, and show
how they can be classified according to their capability to perform some precisely defined
tasks. 
Our knowledge in the subject of entanglement is still far from complete, 
although significant progress has been made in the recent years and 
very active research is currently underway.

\section{Bipartite pure states: Schmidt decomposition}

In this chapter, we will primarily consider bipartite systems, which are traditionally supposed to be in possession of 
Alice (A) and Bob (B),  who can be
located in distant regions. 
Let Alice's physical system be described 
by the Hilbert space \(\mathcal{H}_A\) and that of Bob by \(\mathcal{H}_B\). Then the joint
physical system of Alice and Bob is described by the tensor product Hilbert space 
\(\mathcal{H}_A \otimes \mathcal{H}_{B}\).

\noindent \textbf{Def. 1} \emph{Product and entangled pure states:}\index{Bipartite pure states}\\
\emph{
A pure state, i.e. a projector \(|\psi_{AB}\rangle \langle \psi_{AB}|\)
on a vector $|\psi_{AB}\rangle \in {\cal H}_A \otimes {\cal H}_B $, is a product state if the states of 
local subsystems are also pure states, that is, if  $|\psi_{AB}\rangle= |\psi_A\rangle \otimes |\psi_B\rangle$.
However, there are states that cannot be written in this form. 
These states are called entangled states.}

An example of  entangled state is the well-known singlet state \((\left|01\right\rangle - \left|10\right\rangle)/\sqrt{2}\), where 
\(|0\rangle\) and \(|1\rangle\) are two orthonormal states. 
Operationally, product states correspond  to those states, that can be locally prepared by Alice and Bob 
at two separate locations. 
Entangled states can, however, be prepared only after the particles of Alice and Bob have interacted either directly or
by means of an ancillary system. 
A very useful representation, only valid for pure bipartite states, is the, so-called, Schmidt representation:\\
\textbf{Theorem 1} \emph{Schmidt decomposition:}\index{Schmidt decomposition}\\ 
\emph{Every $|\psi_{AB}\rangle \in {\cal H}_{\cal A} \otimes {\cal H}_{\cal B}$ can be represented in an
appropriately chosen basis as
\begin{equation}
|\psi_{AB}\rangle = \sum _{i=1}^{M} a_i |e_i\rangle \otimes |f_i\rangle,
\end{equation}
where $|e_i\rangle$ ($|f_i\rangle$) form a part of an orthonormal basis in
${\cal H}_{\cal A}$ (${\cal H}_{\cal B}$), $a_i > 0$, and  $\sum_{i=1}^{M} a_i^2=1$, where $M\le dim{\cal H}_{\cal A}, dim{\cal H}_{\cal B}$}.\\
The positive numbers \(a_i\) are known as the Schmidt coefficients of \(|\psi_{AB}\rangle\). Note that product pure states correspond to those
states, whose Schmidt decompositon has one and only one Schmidt coefficient. 
If the decomposition has more than one Schmidt coefficient, the state is entangled. 
Notice that the squares of the Schmidt coefficients of a pure bipartite state \(|\psi_{AB}\rangle\) are the
eigenvalues of either of the reduced density matrices \(\rho_A\) (\(= \mbox{tr}_B \rho_{AB} \)) and 
 \(\rho_B\) (\(= \mbox{tr}_A \rho_{AB}\)) of \(|\psi_{AB}\rangle\).



\section{Bipartite mixed states: Separable and entangled states}

As discussed in the last section, the question whether a given pure bipartite state is separable or entangled is 
straightforward.  One has just to check if the reduced density matrices 
are pure. This condition is equivalent  to the fact that a bipartite pure state has a 
single Schmidt coefficient. The determination of separability for mixed states is much harder, 
and currently lacks a complete answer, even in composite systems of dimension as low as \({\cal C}^2 \otimes {\cal C}^4\).

To reach a formal definition of separable and entangled states, consider the following preparation procedure 
of a bipartite state between Alice and Bob. 
Suppose that Alice prepares her physical system in the state \(|e_{i}\rangle\) and Bob
prepares his physical system in the state \(|f_{i}\rangle\). 
Then, the combined state of their joint physical
system is given by:
\begin{equation}
|e_{i}\rangle \langle e_i |\otimes |f_{i}\rangle \langle f_i|.
\end{equation}
We now assume that they can communicate over a classical channel
(a phone line, for example). 
Then, whenever Alice prepares the state \(|e_{i}\rangle\) (\(i = 1,2, \ldots, K\)), which she does 
with probability \(p_i\), she communicates that to Bob, and correspondingly Bob prepares 
his system in the state \(|f_{i}\rangle\) (\(i = 1,2, \ldots, K\)). Of course, \(\sum_i p_i =1\). The state that they 
prepare is then 
\begin{eqnarray}
\label{eqn:sepmixed}
\rho_{AB} &= \sum_{i=1}^K  p_i |e_i\rangle \langle e_i| \otimes |f_i\rangle \langle f_i|.
\end{eqnarray}
The important point to note here is that the state displayed in Eq. (\ref{eqn:sepmixed}) is the most general 
state that Alice and Bob will be able to prepare by local quantum operations and classical communication (LOCC) \cite{Werner}.

\noindent \textbf{Def. 2} \emph{Separable and entangled mixed states:}\index{Bipartite mixed states}\\
\emph{A mixed state $\rho_{AB}$ is separable if and only if it can be represented as a convex combination of 
the product of projectors on local states as stated   in Eq. 
(\ref{eqn:sepmixed}). Otherwise, the mixed state is said to be entangled.}

Entangled states, therefore, cannot be prepared locally 
by two parties even after communicating over a classical channel. To prepare such states, the physical systems
must be brought together to interact\footnote{Due to the existence of the phenomenon of 
entanglement swapping \cite{ZZHE}, one must suitably enlarge the notion of preparation of entangled 
states. So, an entangled state between two particles can be prepared if and only if, either the two particles 
(call them A and B) themselves come together to interact at a time in the past, or two \emph{other} particles (call them C 
and D) does the same, with C (D) having interacted beforehand with A (B).}. 
Mathematically, a nonlocal unitary operator\footnote{A unitary operator on 
\(\mathcal{H}_{A}\otimes \mathcal{H}_{B}\), is said to be ``nonlocal'', if it is 
not of the form \(U_A \otimes U_B\), with \(U_A\) (\(U_B\)) being a unitary operator acting on 
\(\mathcal{H}_{A}\)
(\(\mathcal{H}_{B}\)).} must \emph{necessarily} act on the physical system described by 
\(\mathcal{H}_{A}\otimes \mathcal{H}_{B}\), to produce an entangled state from an initial separable state.

The question whether a given bipartite state is separable or not turns out to be quite complicated. 
Among the difficulties, we notice that for an arbitrary state \(\rho_{AB}\), there is no stringent 
bound on the value of \(K\) in Eq. (\ref{eqn:sepmixed}), which is only limited by the Caratheodory theorem to  
be \(K \leq (\dim {\cal H} )^2\) with 
${\cal H} = {\cal H}_A \otimes {\cal H}_B$ (see \cite{karnas00}).
Although the general answer to the separability problem still eludes us, there has been significant progress in recent years, and we will review some such directions in the following sections.

\section{Operational entanglement criteria}\index{Entanglement criteria}

In this section, we will introduce some operational entanglement criteria.
In particular, we will discuss the partial transposition criterion \cite{PeresPPT, HorodeckiPPT}, 
and the majorization criterion \cite{NielsenKempe}.
There exist several other criteria 
(see e.g. Refs. \cite{reduction, realignment, Doherty}), which  will not be discussed here. 
However note that, up to now, a necessary and sufficient criterion for detecting 
entanglement  of an arbitrary given mixed state is still lacking.

\subsection{Partial Transposition}\index{Partial transposition}

\noindent \textbf{Def. 3} \emph{Let \(\rho_{AB}\) be a bipartite density matrix, and let us express it as 
\begin{equation}
\rho_{AB} = \sum_{i,j=1}^{N_A} \sum_{\mu,\nu=1}^{N_B} a_{ij}^{\mu\nu} (|i\rangle\langle j|)_A \otimes (|\mu\rangle\langle \nu|)_B,
\end{equation}
where \(\{|i\rangle\}\) (\(i= 1, 2, \ldots, N_A; N_A \leq \dim {\cal H}_A\))  
(\(\{|\mu\rangle\}\) (\(\mu= 1, 2, \ldots, N_B; N_B \leq \dim {\cal H}_B\))) is a set of  orthonormal vectors in \({\cal H}_A\) (\({\cal H}_B\)). 
The partial transposition, \(\rho_{AB}^{T_A}\), of  \(\rho_{AB}\)
with respect to subsytem \(A\), 
is defined as}
\begin{equation}
\label{eq_partial_trans}
\rho_{AB}^{T_A} = 
\sum_{i,j=1}^{N_A} \sum_{\mu,\nu=1}^{N_B} a_{ij}^{\mu\nu} (|j\rangle\langle i|)_A \otimes (|\mu\rangle\langle \nu|)_B.
\end{equation}

A similar definition exists for the partial transposition of \(\rho_{AB}\) with respect to Bob's subsystem. Notice that 
$\rho_{AB}^{T_{B}}\ =(\rho_{AB}^{T_{A}})^{T}$.
Although the partial transposition depends upon the choice of the basis in
which \(\rho_{AB}\) is written, its eigenvalues are basis independent.
We say that a state has  Positive Partial Transposition (PPT) \index{Positive partial transposition},
whenever \(\rho_{AB}^{T_A} \geq 0\), i.e. the eigenvalues of \(\rho_{AB}^{T_{A}}\) are non-negative. 
Otherwise, the state is said to be Non-positive under  Partial Transposition (NPT). 
Note here that transposition is equivalent to time reversal. 

\noindent \textbf{Theorem 2} \cite{PeresPPT}\\
\emph{If  a state $\rho_{AB}$ is separable, then $\rho_{AB}^{T_{A}}\ \ge\ 0$ and
$\rho_{AB}^{T_{B}}\ =\left(\rho_{AB}^{T_{A}}\right)^{T}\ \ge\ 0$.}

\noindent Proof:\\
Since $\rho_{AB}$ is separable, it can be written as
\begin{eqnarray}
\rho_{AB} &
 = \sum_{i=1}^{K}\ p_i |e_i \rangle \langle e_i |\otimes |f_i\rangle \langle f_i | \ge 0.
\end{eqnarray}
Now performing the partial transposition w.r.t. A, we have 
\begin{eqnarray}
\rho_{AB}^{T_{A}} &=& \sum_{i=1}^{K}\ p_i\left(|e_i\rangle \langle e_i |
\right)^{T_{A}}\otimes |f_i \rangle \langle f_i |\nonumber\\
&=& \sum_{i=1}^{K}\ p_i |e_i^*\rangle \langle e_i^*| \otimes | f_i\rangle \langle f_i | \ge 0.
\end{eqnarray}
Note that in the second line, we have used the fact that $A^{\dagger}=\left(A^*\right)^{T}$. \(\square\)

The \emph{partial transposition criterion},
\index{Partial transposition criterion} for detecting entanglement is simple: Given a bipartite 
state \(\rho_{AB}\), find the eigenvalues of any of its partial transpositions. A negative eigenvalue immediately 
implies that the state is entangled. Examples of states for which the partial transposition has negative eigenvalues include
the singlet state. 

The partial transposition criterion allows to detect in a straightforward manner all entangled states that are
NPT states. This is a huge class of states. However, it turns out that there exist 
PPT states which are not separable, as pointed out in Ref. \cite{Pawelbound} (see also \cite{Horodeckibound}). 
Moreover, the set of PPT entangled states is not a set of measure zero
\cite{koto-volume-re}. 
It is, therefore, important to have further independent criteria 
of entanglement detection which permits to detect entangled PPT states. It is worth mentioning that 
PPT states which are entangled, form the only known examples of the ``bound entangled states'' (see Refs. \cite{Horodeckibound,biyog} for details).
Note also that both separable as well as PPT states form convex sets. 

Theorem 2 is a necessary condition of separability in any arbitrary dimension. However, 
for some special cases, the partial transposition criterion is both, a necessary and sufficient condition for
separability:\\
\textbf{Theorem 3} \cite{HorodeckiPPT}\\
\emph{In ${\cal C}^2 \otimes {\cal C}^2$ or ${\cal C}^2 \otimes {\cal C}^3$,
a state $\rho$ is separable if and only if $\rho^{T_{A}}\ \ge\ 0$}.










\subsection{Majorization}\index{Majorization}

The partial transposition criterion, although powerful, is not able to detect entanglement in 
a finite volume of states. It is, therefore, 
interesting to discuss other independent criteria. The majorization criterion\index{Majorization criterion}, to be 
discussed in this subsection, has been recently shown to be \emph{not} more powerful in detecting entanglement. 
We choose to discuss it here, mainly because it has independent roots. Moreover, it reveals a very interesting 
thermodynamical property of entanglement.

Before presenting the criterion,
we must first give the definition of majorization \cite{majorizationBhatia}.\\
\textbf{Def. 3} \emph{Let \(x = (x_1, x_2, \ldots, x_d )\), and \(y  = (y_1, y_2, \ldots, y_d )\) be two probablity distributions, arranged in
decreasing order, i.e. \(x_1 \geq x_2 \geq \ldots \geq x_d\) and \(y_1 \geq y_2 \geq \ldots \geq y_d\). 
Then we define ``\(x\) majorized by \(y\)'', 
denoted as \(x\prec y\), as} 
\begin{equation}
 \sum_{i=1}^l x_i \leq \sum_{i=1}^l y_i,  
 \end{equation}
\emph{where \( l = 1,2, \ldots d-1\), and equality holds when \(l = d\)}.

\noindent \textbf{Theorem 4} \cite{NielsenKempe}\\
\emph{If a state \(\rho_{AB}\) is separable,  then}
\begin{equation}
\label{dibakar}
\lambda (\rho_{AB}) \prec \lambda (\rho_A ), \quad  and \quad \lambda (\rho_{AB}) \prec \lambda (\rho_A ), 
\end{equation}
\emph{where \(\lambda (\rho_{AB})\) is the set of eigenvalues of \( \rho_{AB} \), and 
\(\lambda ( \rho_A )\) and \( \lambda ( \rho_B ) \) are the sets of eigenvalues
of the corresponding reduced density matrix of the state \(\rho_{AB}\), and where all the sets are arranged in decreasing order.}

The majorization criterion: Given a bipartite state, it is entangled if Eq. (\ref{dibakar}) is violated. 
However, it was recently shown in Ref. \cite{Hiroshima}, that a state that is not detected by the 
positive partial transposition criterion, will not be detected by the majorization criterion also. 
Nevertheless, the criterion has other important implications. We will now discuss one such.



Let us reiterate an interesting fact about the singlet state: The global state is pure, while the local states are 
completely mixed. In particular, this implies that the von Neumann entropy\footnote{The von Neumann entropy of a state \(\rho\) is 
\(S(\rho) = -\mbox{tr}\rho \log_2 \rho\).}
 of the global state is lower than either of 
the von Neumann entropies of the local states. The von Neumann entropy can however be used to quantify 
disorder in a quantum state. This implies that there exist bipartite quantum states for which the global disorder can be 
more than either of the local disorders. This is a nonclassical fact as for two classical random variables, the 
Shannon entropy\footnote{The Shannon entropy of a random variable \(X\), taking up values \(X_i\), with probabilities \(p_i\),
is given by \(H(X) = H(\{p_i\}) = - \sum_i p_i \log_2 p_i\).} of the joint distribution cannot be smaller than that of either. 
In Ref. \cite{Horo-mixing}, it was shown that a similar fact is true for separable states:\\
\textbf{Theorem 5}\\
\emph{If a state \(\rho_{AB}\) is separable, 
\begin{equation}
\label{chandrima}
S (\rho_{AB}) \geq S (\rho_A ), \quad  and \quad S (\rho_{AB}) \geq S (\rho_B ). 
\end{equation}
}
Although the von Neumann entropy is an important notion for quantifying disorder, the theory of majorization 
is a more stringent quantifier \cite{majorizationBhatia}:
For two probability distributions \(x\) and \(y\), \(x \prec y\) if and only if \(x =  Dy\), where 
\(D\) is a doubly stochastic matrix\footnote{A matrix \(D= (D_{ij})\) is said to be doubly stochastic, if \(D_{ij}\geq 0\), and 
\(\sum_i D_{ij} = \sum_j D_{ij} =1\).}. Moreover, \(x \prec y\) implies that \(H(\{x_i\}) \geq H(\{y_i\})\).
Quantum mechnics therefore allows the existence of states for which global disoder is greater than local disorder even in the 
sense of majorization.

 A density matrix that satisfies Eq. (\ref{dibakar}), 
automatically satisfies Eq. (\ref{chandrima}). In this sense, Theorem 4 is a generalization of Theorem 5.

\section{Non-operational entanglement criteria}

In this section, we will discuss two further entanglement criteria. We will show how the Hahn-Banach theorem can be used 
to obtain ``entanglement witnesses''. We will also introduce the notion of positive maps, and present the entanglement 
criterion based on it. Both the criteria are ``non-operational'', in the sense that they are not state-independent. 
Nevertheless, they provide important insight into the structure of the set of entangled states. Moreover, the concept of entanglement witnesses
can be used to detect entanglement experimentally, by performing only a few \emph{local} measurements, assuming some prior 
knowledge of the density matrix \cite{sakkhi12, sakkhiprl}.


\subsubsection{Technical Preface}
The following lemma and observation will be useful for later purposes.\\
\textbf{Lemma 1}\\
$\mbox{tr}(\rho_{AB}^{T_{A}}\sigma_{AB}) = \mbox{tr}(\rho_{AB}\sigma_{AB}^{T_{A}})$.

\vspace{0.5cm}
\noindent \textbf{Observation:}\\
The space of linear operators acting on ${\cal H}$ (denoted by ${\cal B}({\cal H})$)
is itself a Hilbert space\index{Hilbert space}, 
with the (Euclidean) scalar product
\begin{equation}
\langle A|B\rangle = \mbox{tr}(A^\dagger B) \qquad A,B \in {\cal B}({\cal H}).
\end{equation}
This scalar product is equivalent to writing $A$ and $B$ row-wise as vectors,
and scalar-multiplying them:
\begin{equation}
\mbox{tr}(A^\dagger B)  = \sum_{ij}A^\ast_{ij}B_{ij} = \sum_{k=1}^{(\dim {\cal H})^2} a_{k}^\ast b_k.
\end{equation}

\subsection{Entanglement Witnesses}

\subsubsection{Entanglement Witness from the Hahn-Banach theorem}

Central to the concept of entanglement witnesses, is the Hahn-Banach
theorem, \index{Hahn-Banach theorem} which we will present here limited to our situation and without
proof (see e.g. \cite{alt:1985} for a proof of the more general theorem):\\
\textbf{Theorem 6}\\
\emph{
Let $S$ be a convex compact set in a finite dimensional Banach space. 
Let $\rho$ be a point in the space with $\rho \not\in S$.
Then there exists a 
hyperplane}\footnote{A hyperplane is a linear subspace with 
dimension one less than the dimension of the space itself.}\index{Hyperplane} \emph{that separates $\rho$ from $S$.}

\begin{figure}[htbp]
\begin{center}
\includegraphics{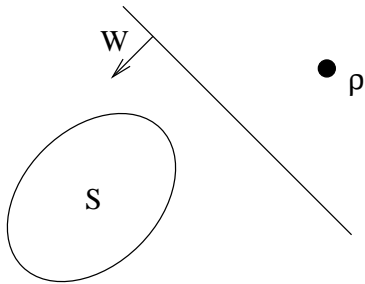}
\end{center}
\caption[Schematic picture of the Hahn-Banach theorem]
{Schematic picture of the Hahn-Banach theorem. The (unique)
unit vector
orthonormal to the hyperplane can be used to define \emph{right} and
\emph{left} in respect to the hyperplane by using the signum of the scalar product.}\label{fig:hyper}
\end{figure}

The statement of the theorem is illustrated in figure \ref{fig:hyper}. The figure
motivates the introduction of a new coordinate system
located within the hyperplane (supplemented by an orthogonal vector $W$ which
is chosen such that it points away from $S$).
Using this coordinate system, every state $\rho$ can be characterized by
its distance from the plane, by projecting $\rho$ onto the chosen orthonormal
vector and using the trace as scalar product, i.e. $\mbox{tr}(W\rho)$. This
measure is either positive, zero, or negative. We now suppose that \(S\) is the convex compact set of all separable states. 
According to our choice of
basis in figure \ref{fig:hyper}, every separable state has a positive
distance while there are some entangled states with a negative distance.
More formally this can be phrased as:\\
\textbf{Def. 4}
\emph{A hermitian operator (an observable) $W$ is called an entanglement
witness (EW)\index{Entanglement witness (EW)} if and only if}
\begin{equation}
\exists \rho \quad \mbox{such that} \quad \mbox{tr}(W \rho) < 0, \qquad
\mbox{while} \quad \forall \sigma \in S, \quad \mbox{tr}(W\sigma) \geq 0.
\end{equation}


\noindent \textbf{Def. 5}\index{Decomposable entanglement witness}
\emph{An EW is decomposable if and only if there exists operators $P$, $Q$ with}
\begin{equation}
W = P + Q^{T_{A}}, \qquad P,Q \geq 0.
\end{equation}

\noindent \textbf{Lemma 2}\\
\emph{Decomposable EW cannot detect PPT entangled states.}

\noindent Proof:\\
Let $\delta$ be a PPT entangled state and $W$ be a decomposable EW.  Then
\begin{equation}\label{eqn:dEWnd}
\mbox{tr}(W\delta) = \mbox{tr}(P\delta) + \mbox{tr}(Q^{T_{A}} \delta)
= \mbox{tr}(P\delta)+\mbox{tr}(Q\delta^{T_{A}}) \geq 0.
\end{equation}
Here we used Lemma 1.\(\square\)

\noindent \textbf{Def. 6}
\index{Non-decomposable entanglement witness (nd-EW)}
\emph{An EW is called non-decomposable entanglement witness (nd-EW) if and only if
there exists at least one PPT entangled state which is detected by the witness.}

Using these definitions, we can restate the consequences of the
Hahn-Banach theorem in several ways:\\
\textbf{Theorem 7} \cite{Woronowicz,HorodeckiPPT, Terhal, sakkhi-ager}
\begin{enumerate}
\item $\rho$ \emph{is entangled if and only if,  $\exists$ a witness $W$, such that} $\mbox{tr}(\rho W) < 0$.
\item $\rho$ \emph{is a PPT entangled state  if and only if  $\exists$ a nd-EW, $W$, such that} $\mbox{tr}(\rho W) < 0$.
\item $\sigma$ \emph{is separable  if and only if  $\forall$ EW,} $\mbox{tr}(W\sigma) \geq 0$.
\end{enumerate}
From a theoretical point of view, the theorem is quite powerful.
However, it does not give any insight of how to construct for a given state $\rho$, the appropriate witnes operator.

\begin{figure}
\begin{center}
\includegraphics{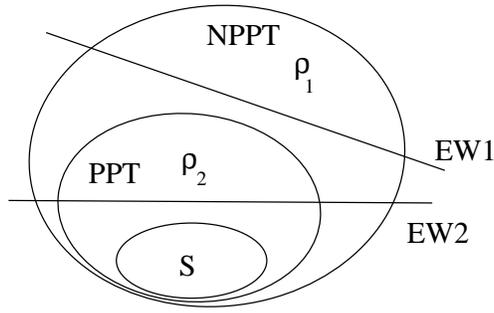}
\end{center}
\caption[Comparison of a nondecomposable entanglement witness with a decomposable one]
{Schematic view of the Hilbert-space with two states $\rho_1$ and
$\rho_2$ and two witnesses $EW1$ and $EW2$. $EW1$ is a decomposable EW, and it
 detects only  NPT states like $\rho_1$. $EW2$ is an  nd-EW, and it
detects also some PPT states like $\rho_2$. Note that neither witness
detects \emph{all} entangled states.}\label{fig:schhil}
\end{figure}
\subsubsection{Examples}\label{sec:E}
For a  decomposable witness
\begin{equation}
W' = P + Q^{T_{A}},
\end{equation}
\begin{equation}
\label{narayani}
\mbox{tr}(W'\sigma) \geq 0,
\end{equation}
for all separable states \(\sigma\).

\noindent Proof:\\
If $\sigma$ is separable, then it can be written as a convex sum of product
vectors.
So if Eq. (\ref{narayani}) holds for any product vector \(|e,f\rangle\),
any separable state will  also satisfy the same.
\begin{eqnarray}
\mbox{tr}(W' |e,f\rangle \langle e,f|) &=&
\langle e,f |W' |e,f\rangle \nonumber \\
&=&
\underbrace{\langle e,f | P |e,f \rangle}_{\geq 0} + \underbrace{\langle e,f | Q^{T_{A}} |e,f \rangle }_{\geq 0} , 
\end{eqnarray}
because
\begin{equation}
\langle e,f | Q^{T_{A}} |e,f \rangle = \mbox{tr} (Q^{T_{A}} |e,f\rangle \langle e,f |) = \mbox{tr} (Q |e^\ast,f \rangle \langle e^\ast,f |  ) 
\geq 0.
\end{equation}
Here we used Lemma 1, and $P,Q \geq 0$. \(\square\)
This argumentation shows that $W=Q^{T_{A}}$ is a suitable witness also.

Let us consider 
the simplest case of ${\cal C}^2 \otimes {\cal C}^2$.  We can use
\begin{equation}
|\phi^+ \rangle = \frac{1}{\sqrt{2}}\left(|00\rangle + |11\rangle\right),
\end{equation}
to write the density matrix
\begin{equation}
Q = \left( \begin{array}{cccc}
\frac12 & 0 & 0 & \frac12 \\
0 & 0 & 0 & 0 \\
0 & 0 & 0 & 0 \\
\frac12 & 0 & 0 & \frac12 \end{array}\right).  \quad \mbox{Then} \quad
Q^{T_{A}} = \left( \begin{array}{cccc}
\frac12 & 0 & 0 & 0 \\
0 & 0 & \frac12 & 0 \\
0 & \frac12 & 0 & 0 \\
0 & 0 & 0 & \frac12 \end{array}\right).
\end{equation}
One can quickly verify that indeed $W=Q^{T_{A}}$ fulfills the witness
requirements. Using
\begin{equation}
|\psi^- \rangle = \frac{1}{\sqrt{2}} \left(|01 \rangle - |10 \rangle\right),
\end{equation}
we can rewrite the witness:
\begin{equation}\label{eqn:exwit}
W = Q^{T_{A}} = \frac{1}{2} \left(I - 2 |\psi^- \rangle \langle \psi^- |\right).
\end{equation}
This witness now detects $|\psi^- \rangle$:
\begin{equation}
\mbox{tr} (W |\psi^- \rangle \langle \psi^- |) = -\frac{1}{2}.
\end{equation}


\subsection{Positive maps}\index{Positive maps (PM)}

\subsubsection{Introduction and definitions}
So far we have only considered states 
belonging to a Hilbert space ${\cal H}$, and operators acting on the
Hilbert space. However, the space of operators ${\cal B}({\cal H})$ has also a Hilbert space structure. 
We now look at transformations of operators, the so-called maps\index{Map} which can be regarded as 
\emph{superoperators}\index{Operator>Super}. As we will see, this will lead us to an important 
characterization of entangled and separable states.
We start by defining linear maps:\\
\textbf{Def. 7}
\emph{A linear, self-adjoint map $\epsilon$ is a transformation} \index{Linear map} \index{Self-adjoint map}
\begin{eqnarray}
\epsilon: {\cal B} ({\cal H}_{B}) \rightarrow {\cal B} ({\cal H}_{C}),
\end{eqnarray}
\emph{which}
\begin{itemize}
\item \emph{is linear, i.e.}
\begin{eqnarray}
\epsilon(\alpha O_1+\beta O_2) &= \alpha \epsilon(O_1) + 
\beta \epsilon(O_2)\quad\forall O_1,\,O_2\in {\cal B}({\cal H}_{B}),
\end{eqnarray}
\emph{where \(\alpha\), \(\beta\) are complex numbers,}
\item \emph{and maps hermitian operators onto hermitian operators, i.e.}
\begin{eqnarray}
\epsilon(O^{\dagger})&=\left(\epsilon(O)\right)^{\dagger}\qquad\forall O \in {\cal B}({\cal H}_{B}).
\end{eqnarray}
\end{itemize}
For brevity, we will only write ``linear map'', instead of ``linear self adjoint map''. 
The following definitions help to further characterize linear maps.\\
\textbf{Def. 8} \emph{A linear map $\epsilon$ is called trace preserving if }
\begin{eqnarray}
\mbox{tr}(\epsilon(O))=\mbox{tr}(O)\quad\forall O\in {\cal B}({\cal H}_{B}).
\end{eqnarray}

\noindent \textbf{Def. 9} \emph{Positive map:}\\
\emph{A linear, self-adjoint map $\epsilon$ is called positive if}
\begin{eqnarray}
\forall \rho \in {\cal B}({\cal H}_{B}), \quad
\rho \geq 0 \quad \Rightarrow \quad
\epsilon(\rho) \geq 0.
\end{eqnarray}
Positive maps have, therefore, the property of mapping positive operators onto
positive operators. It turns out that by considering maps that are a tensor product of
a positive operator acting on subsystem A, and the identity acting  on subsystem
B, one can learn about the properties of the composite system.  

\noindent \textbf{Def. 10} \emph{Completely positive map:} \index{Completely positive map}\\
\emph{A positive linear map $\epsilon$ is completely positive if for any tensor extension
of the form}
\begin{equation}
\epsilon'  = {\cal I}_A\otimes\epsilon,
\end{equation}
where 
\begin{equation}
\epsilon' :        {\cal B}({\cal H}_{A} \otimes {\cal H}_{B}) 
   \rightarrow  {\cal B}({\cal H}_{A} \otimes {\cal H}_{C}), 
\end{equation}
\emph{$\epsilon'$ is positive. Here \({\cal I}_A\) is the identity map on \({\cal B}({\cal H}_{A})\).}

\textbf{Example: Hamiltonian evolution of a quantum system.}
Let $O\in {\cal B}({\cal H}_B)$ and $U$ an unitary matrix and let us define $\epsilon$ by
\begin{eqnarray}
\epsilon:  {\cal B}( {\cal H}_A) & \rightarrow & {\cal B}( {\cal H}_A) \nonumber \\
 \epsilon(O) & = & UOU^{\dagger}.
\end{eqnarray}
As an example for this map, consider the time-evolution of a density matrix. It
can be written as $\rho(t)=U(t)\rho(0)U^{\dagger}(t)$, i.e. in the form given above. 
Clearly this map is linear, self-adjoint, positive
and trace-preserving. It is also completely positive, because for 
$0\leq w\in {\cal B}({\cal H}_A \otimes {\cal H}_A)$,
\begin{equation}
({\cal I}_A\otimes\epsilon)w  = (I_A\otimes U)w(I_A\otimes U^{\dagger})
=\tilde{U}w\tilde{U}^{\dagger},
\end{equation}
where $\tilde{U}$ is unitary. But then $\langle \psi |\tilde{U}w\tilde{U}^{\dagger} |\psi \rangle \geq 0$,
if and only if $\langle \psi | w |\psi \rangle \geq 0$ (since positivity is not changed by unitary evolution).

\textbf{Example: Transposition.}
An example of a positive but not completely positive map is the transposition $T$ defined as:
\begin{eqnarray}
T:  {\cal B}({\cal H}_B) & \rightarrow & {\cal B}({\cal H}_B) \nonumber \\
 T(\rho) & = & \rho^{T}.
\end{eqnarray}
Of course this map is positive, but it is not completely positive, because
\begin{equation}
({\cal I}_A\otimes T)w=w^{T_{B}},
\end{equation}
and we know that there exist states for which $\rho\geq 0$, but $\rho^{T_{B}}\not\geq 0$.

\noindent \textbf{Def. 11}
\emph{A positive map is called decomposable if and only if it can be written as}
\begin{equation}
\epsilon=\epsilon_1+\epsilon_2 T
\end{equation}
\emph{where $\epsilon_1$, $\epsilon_2$ are completely positive maps and $T$ is the 
operation of transposition.}

\subsubsection{Positive maps and entangled states}

Partial transposition can be regarded as a particular case of a map that is positive but not completely positive. 
We have already seen that this particular positive but not completely positive map gives us a 
way to discriminate entangled states from separable states.  
The theory of positive maps provides with stonger conditions for separability, as shown in Ref. \cite{HorodeckiPPT}.\\
\textbf{Theorem 8}\\
\emph{A state $\rho\in {\cal B} ({\cal H}_A \otimes {\cal H}_B)$ is separable if and only if for all positive maps}
\begin{equation}
\epsilon:  {\cal B} ({\cal H}_B)  \rightarrow {\cal B} ({\cal H}_C),
\end{equation}
\emph{we have}
\begin{equation}
({\cal I}_A\otimes\epsilon)\rho\geq 0.
\end{equation}

\noindent Proof: 

[$\Rightarrow$] As $\rho$ is separable,   we can write it as
\begin{equation}
\rho
=\sum_{k=1}^P p_k |e_k \rangle \langle e_k | \otimes |f_k \rangle \langle f_k |,
\end{equation}
for some $P>0$. On this state, $({\cal I}_A\otimes\epsilon)$ acts as
\begin{equation}
({\cal I}_A\otimes\epsilon)\rho 
=\sum_{k=1}^P p_k |e_k \rangle \langle e_k | \otimes \epsilon \left(|f_k \rangle \langle f_k |\right)
\geq 0,
\end{equation}
where the last $\geq$ follows because $|f_k \rangle \langle f_k |\geq0$, and $\epsilon$ is positive.\\
$[\Leftarrow]$ The proof in this direction is not as easy as the only if direction. We shall prove it 
at the end of this section.\\
Theorem 8 can also be recasted into the following form:\\
\textbf{Theorem 8} \cite{HorodeckiPPT}\\
\emph{A state $\rho\in {\cal B} ({\cal H}_A \otimes {\cal H}_B)$  is entangled if and only if there exists a positive
map \(\epsilon:  {\cal B} ({\cal H}_B)  \rightarrow {\cal B} ({\cal H}_C)\), such that}
\begin{equation}
\label{chot-bogoley}
({\cal I}_A\otimes\epsilon)\rho\not\geq 0.
\end{equation}
Note that Eq. (\ref{chot-bogoley}) can never hold for maps, \(\epsilon\), that are completely positive, and for 
non-positive maps, it may hold even for separable states. Hence, any positive but not completely positive map 
can be used to detect entanglement. 

\subsubsection{Jamio{\l}kowski Isomorphism}
In order to complete the proof of Theorem 8,
we introduce first the 
Jamio{\l}kowski isomorphism \cite{jamiolkowski}
\index{Jamio{\l}kowski isomorphism}
between operators and maps.
Given an operator $E\in {\cal B} ({\cal H}_B \otimes {\cal H}_C)$, and an orthonormal product basis $|k,l \rangle$,
we define a map by
\begin{eqnarray}
\epsilon: {\cal B} ({\cal H}_B)  &\rightarrow & {\cal B} ({\cal H}_C) \nonumber \\
\epsilon(\rho) & = & \sum_{k_1,l_1,k_2,l_2}\,{}_{BC}\langle k_1l_1| E | k_2l_2 \rangle_{BC} \quad
|l_1\rangle_{CB} \langle k_1 | \rho |k_2\rangle_{BC} \langle l_2|,
\end{eqnarray}
or in short form,
\begin{equation}
\epsilon(\rho)= \mbox{tr}_{B}(E\rho^{T{B}}).
\end{equation}
This shows how to construct the map $\epsilon$ from a given operator $E$. To construct an
operator from a given map we use the state
\begin{equation}
|\psi^+ \rangle =\frac{1}{\sqrt{M}}\sum_{i=1}^{M} |i\rangle_{B'} |i\rangle_{B}
\end{equation}
(where $M=\dim {\cal H}_B$) to get
\begin{equation}
M\left(I_{B'}\otimes\epsilon\right)\left( |\psi^+\rangle \langle \psi^+| \right)=E.
\end{equation}
This isomorphism between maps and operators results in the following properties:\\
\textbf{Theorem 10} \cite{jamiolkowski, Woronowicz,HorodeckiPPT, Terhal, sakkhi-ager}\\
\emph{
\begin{enumerate}
\item $E\geq0$ if and only if $\epsilon$ is a completely positive map.
\item $E$ is an entanglement witness if and only if $\epsilon$ is a positive map.
\item $E$ is a decomposable entanglement witness if and only if $\epsilon$ is decomposable.
\item $E$ is a non-decomposable entanglement witness if and only if $\epsilon$ is non-decomposable and positive.
\end{enumerate}
}
To indicate further how this equivalence between maps and opertors works, we develop here a proof for  the ``only if" direction of the second statement. Let
$E \in {\cal B}({\cal H}_B \otimes {\cal H}_C)$ be an entanglement witness, then $\langle e,f| E |e,f \rangle \geq 0$.
By the Jamio{\l}kowski isomorphism, the corresponding map is defined as 
$\epsilon(\rho)=\mbox{tr}_{B}(E\rho^{T_{B}})$ where $\rho \in {\cal B}({\cal H}_B)$.
We have to show that
\begin{equation}
{}_C \langle \phi | \epsilon(\rho) |\phi\rangle_C={}_C \langle \phi| \mbox{tr}(E\rho^{T_{B}}) |\phi \rangle_C \geq 0\qquad 
\forall |\phi\rangle_C \in {\cal H}_C.
\end{equation}
Since $\rho$ acts on Bob's space, using the spectral decomposition of $\rho$,
\(\rho=\sum_i\lambda_i |\psi_i \rangle \langle \psi_i \), leads to 
\begin{equation}
\rho^{T_{B}}=\sum_i\lambda_i 
|\psi_i^\ast \rangle \langle \psi_i^\ast |,
\end{equation}  
where all $\lambda_i\ge 0$. Then
\begin{eqnarray}
{}_C\langle\phi | \epsilon(\rho)|\phi\rangle_C  &=&
{}_C\langle\phi | \sum_i \mbox{tr}_B (E\lambda_i |\psi_i^\ast\rangle_B {}_B \langle \psi_i^\ast |) |\phi\rangle_C \nonumber\\
&=& \sum_i\lambda_i {}_{BC} \langle \psi_i^\ast,\phi | E |\psi_i^\ast,\phi \rangle_{BC} \geq 0.
\end{eqnarray}
\(\square\)

We can now proof the $\Leftarrow$ direction of Theorem 8
or, equivalently,
the $\Rightarrow$ direction of Theorem 9.
We thus have to show that if $\rho_{AB}$
is entangled, there exists a positive map $\epsilon: {\cal B}({\cal H}_A) \rightarrow {\cal B}({\cal H}_C)$,
such that $\left(\epsilon\otimes {\cal I}_B\right)\rho$ is not positive definite.
If $\rho$ is entangled, then there exists an entanglement witness
$W_{AB}$ such that
\begin{eqnarray}
\mbox{tr}(W_{AB}\rho_{AB}) < 0, \quad \mbox{and} \nonumber \\
\mbox{tr}(W_{AB}\sigma_{AB}) \geq 0,
\end{eqnarray}
for all separable $\sigma_{AB}$. $W_{AB}$ is an entanglement witness (which detects
$\rho_{AB}$) if and only if $W_{AB}^{T}$ (note the complete transposition!) is also an entanglement witness (which detects
$\rho_{AB}^{T}$).
%
We define a map by
\begin{eqnarray}
\epsilon: {\cal B}({\cal H}_A) & \rightarrow & {\cal B}({\cal H}_C),\\
 \epsilon(\rho) & = & \mbox{tr}_{A} (W^{T}_{AC}\rho^{T_{A}}_{AB}),
\end{eqnarray}
where $\dim {\cal H}_C=\dim {\cal H}_B = M$. Then
\begin{equation}
(\epsilon\otimes {\cal I}_B)(\rho_{AB})= \mbox{tr}_{A}(W_{AC}^{T}\rho_{AB}^{T_{A}})=\mbox{tr}_{A}(W_{AC}^{T_{C}}\rho_{AB})=
\tilde{\rho}_{CB},
\end{equation}
where we have used Lemma 1,
and that $T=T_{A} \circ T_{C}$.
To complete the proof, one has to show that  $\tilde{\rho}_{CB}\not\ge 0$, which can be done by showing that 
\({}_{CB}\langle \psi^+| \tilde{\rho}_{CB} |\psi^+\rangle_{CB} < 0\), where 
\(|\psi^+ \rangle_{CB}=\frac{1}{\sqrt{M}} \sum_i |ii\rangle_{CB}\), with \(\{|i\rangle\}\) being  an orthonormal basis. 
\(\square\)

\section{Bell inequalities}

The first criterion used to detect entanglement was  Bell inequalities, which we briefly discuss in this section. 
As we shall see, Bell inequalities are essentially a special type of entanglement witness. 
An additional property of Bell inequalities is that any entangled state detected by them is nonclassical 
in a particular way: It violates ``local realism''. 

The assumptions of ``locality'' and ``realism'' were already present in the famous argument of Einstein, Podolsky, and Rosen
\cite{EPRparadox}, that questions the completeness of quantum mechanics. 
%
%
%
%
Bell \cite{Bell} made these assumptions more precise, and more importantly,
showed that the assumptions are actually \emph{testable} in experiments. He
derived an inequality that must be satisfied by any physical theory of nature, that is ``local''
as well as ``realistic'', the precise meanings of which will be described below. The inequality is 
actually a constraint on a linear function of results of certain experiments. He then went on to show
that there exist states in quantum theory that violate this inequality in experiments. 
Modulo some so-called loopholes (see e.g. \cite{gerakol}), these inequalities have been shown to be actually violated in
 experiments (see e.g. \cite{Paris-ghyama} and references therein). 
In this section, we will first derive a Bell  inequality\footnote{We do not derive here the original 
Bell inequality, which Bell derived in 1964 \cite{Bell}. Instead, we  derive the stronger form of the Bell 
inequality which Clauser, Horne,  Shimony, and Holt (CHSH) derived in 1969 \cite{CHSHineq}. 
A similar  derivation was also given by 
Bell himself in 1971 \cite{Bell71}.}
and then show how this inequality 
is violated by the singlet state.

Consider a two spin-1/2 particle state where the two particles are far apart. Let the particles
be called
\(A\) and \(B\). Let projection valued  measurements in the directions
 \(a\) and \(b\) be done on \(A\) and \(B\) respectively.  The 
outcomes of the measurements performed on the particles \(A\) and \(B\) in the 
directions \(a\) and \(b\), are respectively \(A_a\) and \(B_b\). The measurement result 
\(A_a\) (\(B_b\)), whose values can be \(\pm 1\),
 may depend on the direction \(a\) (\(b\))
  and some other uncontrolled parameter \(\lambda\) which may depend on anything, that is, 
may depend upon system or measuring device or both. Therefore we assume that \(A_a\) (\(B_b\)) 
has a definite pre-measurement value \(A_a (\lambda)\) (\(B_b (\lambda)\)). 
Measurement merely uncovers this value. This is the \emph{assumption of reality}.  \(\lambda\) is
usually called a hidden variable and this assumption is also termed as the hidden variable assumption. 
Moreover, the measurement result at A (B) does not depend on what measurements are performed at B (A). That is,
for example \(A_a(\lambda)\) does not depend upon \(b\). This is the \emph{assumption of locality},
also called the Einstein's locality  assumption.
The parameter \(\lambda\) 
is
assumed to have 
 a probability distribution, say \(\rho(\lambda)\). Therefore
 \(\rho(\lambda)\) satisfies the following:
\begin{equation}
\int \rho(\lambda) d\lambda = 1, \quad \rho(\lambda) \geq 0.
\end{equation}
The correlation  function of the two spin-1/2 particle state 
for a measurement in a fixed direction \(a\) for particle \(A\) and 
\(b\) for particle \(B\), is then given by (provided the hidden variables exist)
\begin{equation}
\label{eq_correlation}
E(a, b) = \int A_a(\lambda) B_b(\lambda) \rho(\lambda) d\lambda.
\end{equation}
Here 
\begin{equation}
A_a(\lambda) = \pm 1, \quad  \mbox{and} \quad  B_b(\lambda) = \pm 1,
\end{equation}
because the measurement values were assumed to be \(\pm 1\).

Let us now suppose that the observers at the  
two particles \(A\) and \(B\) can choose their measurements 
from two observables \(a\), \(a^{'}\) and \(b\), \(b^{'}\) respectively,
and the corresponding 
outcomes are \(A_a\), \(A_{a^{'}}\) and \(B_b\), \(B_{b^{'}}\) respectively. Then
\begin{eqnarray}
E(a, b) + E(a, b^{'}) + E(a^{'}, b) - E(a^{'}, b^{'}) \nonumber \\ = 
\int [A_a (\lambda)(B_b(\lambda)+ B_{b^{'}}( \lambda)) + 
       A_{a^{'}} (\lambda)(B_b(\lambda)- B_{b^{'}}( \lambda)) ]
                \rho(\lambda) d\lambda .
\end{eqnarray}
Now \(B_b(\lambda)+ B_{b^{'}}( \lambda)\)  and \(B_b(\lambda)- B_{b^{'}}( \lambda)\) can only
be \(\pm 2\) and \(0\), or  \(0\) and \(\pm 2\) respectively. Consequently,
\begin{equation}
\label{eq_Bell_inequality}
-2 \leq E(a, b) + E(a, b^{'}) +  E(a^{'}, b) - E(a^{'}, b^{'})  \leq 2.
\end{equation}
This is the well-known CHSH inequality.  Note here that in obtaining the above inequality,
we have never used quantum mechanics. We have only assumed Einstein's locality principle and
an underlying hidden variable model.
Consequently, a Bell inequality is a constraint that any physical theory that is both, local and realistic, has to
satisfy. 
Below,  we will show that this inequality can be violated
by a quantum state.  Hence quantum mechanics is incompatible with an underlying local realistic model.

\subsection{Detection of entanglement by Bell inequality}
Let us now show how the singlet state can be detected by a Bell inequality. 
This additionally will indicate that quantum theory is incompatible with local realism.
For the singlet state, the quantum mechanical prediction of the 
correlation function \(E(a, b)\) is given by 
 \begin{equation}
 \label{pukurchuri}
E(a, b) = \left \langle \psi^{-} | \sigma_a \cdot \sigma_b| \psi^{-} \right\rangle = - \cos(\theta_{ab}),
\end{equation} 
where \(\sigma_a = \vec{\sigma} \cdot \vec{a} \) and similarly for \(\sigma_b\). 
\(\vec{\sigma}= (\sigma_x, \sigma_y, \sigma_z)\), where \(\sigma_x\), \(\sigma_y\) and  \(\sigma_z\) are
the Pauli spin matrices.
And  \(\theta_{ab}\) is the angle 
between the two measurement directions \(a\) and \(b\).

\begin{figure}[ht]
\begin{center}
\unitlength=0.5mm
\begin{picture}(80,50)(0,-10)
\thicklines
\put(10,15){\line(1,0){60}}
\put(40,48){\(a\)}
\put(70,45){\(b\)}

\put(70,15){\(a^{'}\)}
\put(70,-15){\(b^{'}\)}

\put(40,-15){\line(0,1){60}}
\put(15,-10){\line(1,1){50}}
\put(15,40){\line(1,-1){50}}

\put(45, 30){\(\frac{\pi}{4}\)}

\end{picture}
\end{center}
\caption[Optimal apparatus setting for violation of Bell-CHSH inequality by singlet.]{Schematic diagram 
showing the direction of \(a\), \(b\), \(a^{'}\), \(b^{'}\)
for obtaining maximal violation of Bell inequality by the singlet state.}
\label{fig_direction_highest} 
\end{figure}
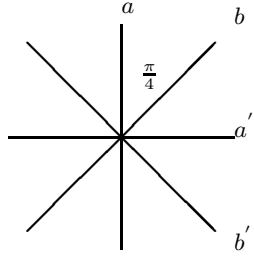

So for the singlet state, one has 
\begin{eqnarray}
\label{khalchuri}
B_{CHSH}&= & E(a, b) + E(a, b^{'}) +  E(a^{'}, b) - E(a^{'}, b^{'}) \nonumber \\
&= & - \cos \theta_{ab} - \cos \theta_{a b^{'}} - \cos \theta_{a^{'}b} + \cos \theta_{a^{'}b^{'}}.
\end{eqnarray}
The maximum value of this function is attained for the directions \(a\), \(b\), \(a^{'}\), \(b^{'}\)
on a plane, as given in Fig. \ref{fig_direction_highest}, and in that case 
\begin{equation}
|B_{CHSH}| = 2 \sqrt{2}.
\end{equation}

This clearly violates the inequality in Eq. (\ref{eq_Bell_inequality}). But Eq. 
(\ref{eq_Bell_inequality})
was a constraint 
for any physical theory which 
has an underlying local hidden variable model.  As the singlet state,
a state allowed by the quantum mechanical description of nature, violates the constraint (\ref{eq_Bell_inequality}),
quantum mechanics cannot have an underlying local hidden variable model. In other words, quantum mechanics is not 
local realistic. This is the statement of the celebrated Bell theorem.

Moreover, it is easy to convince oneself that any separable state does have a local realistic description, so that 
such a state cannot violate a Bell inequality. Consequently, the violation of Bell inequality by the singlet state indicates 
that the singlet state is an entangled state. Further, the operator (cf. Eqs. (\ref{pukurchuri}) and (\ref{khalchuri})) 
\begin{equation}
\tilde{B}_{CHSH} = \sigma_a \cdot \sigma_b + \sigma_a \cdot \sigma_{b^{'}} + \sigma_{a^{'}} \cdot \sigma_b 
- \sigma_{a^{'}} \cdot \sigma_{b^{'}} 
\end{equation}
can, by suitable scaling and change of origin, be considered as an entanglement witness for the singlet state, for 
\(a\), \(b\), \(a^{'}\), \(b^{'}\) chosen as in 
figure \ref{fig_direction_highest} (cf. \cite{POMD}).

\section{Classification of bipartite states with respect to quantum dense coding}
Up to now, we have been interested in splitting the set of all bipartite quantum states into separable and entangled states. 
However, one of the main motivations behind the study of entangled states is that some of them can be used to perform 
certain tasks, which are not possible if one uses states without entanglement. It is, therefore, important 
to find out which entangled states are useful for a given task. We discuss here the particular example of quantum dense coding \cite{BW}.

Suppose that Alice wants to send two bits of 
classical information to Bob.  Then a general result known as the Holevo bound (to be discussed below), shows that 
Alice must send two qubits (i.e. 2 two-dimensional quantum systems) to Bob, if only a noiseless quantum channel 
is available. However, if additionally Alice and Bob have previously shared entanglement,
then Alice may have to send less than two qubits to Bob. It was shown by Bennett and Wiesner \cite{BW}, that 
by using a previously shared singlet (between Alice and Bob), Alice will be able to send two bits to Bob, by sending just a 
single qubit.

The protocol of dense coding \cite{BW} works as follows. Assume that Alice and Bob share a singlet state 
\begin{equation}
|\psi^-\rangle = \frac{1}{\sqrt{2}}\left(|01\rangle - |10\rangle\right).
\end{equation}
The crucial observation is that this entangled two-qubit state can be transformed into 
four orthogonal states of the two-qubit Hilbert space by performing unitary operations on just a single qubit. 
For instance, Alice can apply a rotation (the Pauli operations)  or do nothing to her part of the singlet, while Bob does nothing, to obtain the three triplets (or the singlet):
\begin{eqnarray}
\sigma_x \otimes I |\psi^-\rangle = - |\phi^-\rangle, & \quad & \sigma_y \otimes  I |\psi^-\rangle = i |\phi^+\rangle, \nonumber\\
\sigma_z \otimes I |\psi^-\rangle =  |\psi^+\rangle,  & \quad  & I \otimes I |\psi^-\rangle      =  |\psi^-\rangle, 
\end{eqnarray}
where
\begin{eqnarray}
\label{gorur-garir-headlight}
|\psi^\pm\rangle &=& \frac{1}{\sqrt{2}}\left(|01\rangle \pm |10\rangle\right), \nonumber\\
|\phi^\pm\rangle &=& \frac{1}{\sqrt{2}}\left(|00\rangle \pm |11\rangle\right), 
\end{eqnarray}
and \(I\) is the qubit identity operator.  Suppose that the classical information that Alice wants to send to Bob is \(i\), where 
\(i= 0,1,2,3\). Alice and Bob previously agree on the following correspondence between the operations applied at Alice's end and 
the information \(i\) that she wants to send:
\begin{eqnarray}
\sigma_x \Rightarrow i=0, & \quad & \sigma_y \Rightarrow i=1, \nonumber\\
\sigma_z \Rightarrow i=2, & \quad & I \Rightarrow i=3.
\end{eqnarray} 
Depending on the classical information she wishes to send, Alice applies the appropriate rotation on her part of the shared singlet, according 
to the above correspondence. Afterwards, Alice sends her part of the shared state to Bob, via the noiseless 
quantum channel. Bob now has in his possession, the entire two-qubit state, which is in any of the four Bell states 
\(\left\{|\psi^\pm\rangle, |\phi^\pm\rangle \right\}\). Since these states are mutually orthogonal, he will 
be able to distinguish between them and hence find out the classical information sent by Alice.

To consider a more realistic scenario, usually two avenues are taken. One approach 
is to consider a \emph{noisy} quantum channel, while the additional resource is an arbitrary amount of shared bipartite \emph{pure}
state entanglement (see e.g. \cite{Bennett-ek, Bennett-dui}, see also \cite{MarieCurie, Debu}). 
The other approach
is to consider a \emph{noiseless} quantum channel, while the assistance is by a 
given bipartite \emph{mixed} entangled state (see e.g. \cite{MarieCurie,Debu,ek1,dui1, char1, tin1}).

Here, we consider the second approach, 
and derive 
the capacity of dense coding 
in this scenario, for a given state, where the the capacity is defined as the number of classical bits that can be
accessed 
by the receivers, per usage of the noiseless channel.
This will lead to a 
classification of bipartite states according to their degree of ability to 
assist in dense coding. 
In the case where a noisy channel and an arbitrary amount of shared pure entanglement is considered, 
the capacity refers to the channel (see e.g. \cite{Bennett-ek, Bennett-dui}). However, in our case when a noiseless channel and a 
given shared (possibly mixed) state is considered, the capacity refers to the state. Note that the mixed shared state in our case can be thought 
of as an output of a noisy channel. 
A crucial element in finding the capacity of dense coding 
is the Holevo bound \cite{ref-halum}, which 
is a universal upper bound on classical information that can be decoded from a quantum ensemble. Below we  
discuss the bound, 
and subsequently derive the capacity of dense coding.

\subsection{The Holevo bound}
\label{sec-halum}

The Holevo bound is an upper bound on the amount of classical information that can be accessed from a quantum ensemble in which the information is 
encoded. 
Suppose therefore that Alice (\(A\)) obtains the classical message \(i\) that occurs with probability \(p_i\), 
and she wants to send it to Bob (\(B\)). Alice encodes this information \(i\) in a quantum state \(\rho_i\), and sends it to Bob. 
Bob receives the ensemble \(\{p_i, \rho_i\}\), and wants to obtain as much information as possible about \(i\). To do so, 
he performs a measurement, that gives the result \(m\), with probability \(q_m\). Let the corresponding post-measurement ensemble be 
\(\{p_{i|m}, \rho_{i|m}\}\). The information gathered can be quantified by the mutual information between the 
message index \(i\) and the measurement outcome \cite{Chennai}:
\begin{equation}
I(i:m)= H(\{p_i\}) - \sum_m q_m H(\{p_{i|m}\}).
\end{equation}
Note that the mutual information can be seen as the difference between the initial disorder and the (average) final disorder. 
Bob will be interested to obtain the maximal information, which is maximum of \(I(i:m)\) for all measurement strategies. This 
quantity is called the accessible information:
\begin{equation}
I_{acc} = \max I(i:m),
\end{equation} 
where the maximization is over all measurement strategies.

The maximization involved in the definition of accessible information is usually hard to compute, and hence the importance of bounds 
\cite{ref-halum,Utpakhi}. In particular, in Ref. \cite{ref-halum}, a universal upper bound, the Holevo bound, on \(I_{acc}\) is 
given: 
\begin{equation}
I_{acc}(\{p_i, \rho_i\}) \leq \chi(\{p_i, \rho_i\}) \equiv S(\overline{\rho}) - \sum_i p_i S(\rho_i).
\end{equation}
See also \cite{Rajabazar1,Khajuraho,Rajabazar2}. Here \(\overline{\rho} = \sum_ip_i\rho_i\) is the average ensemble state, and 
\(S(\varsigma)= - \mbox{tr}(\varsigma \log_2 \varsigma)\) 
is the von Neumann entropy of \(\varsigma\).

The Holevo bound is asymptotically  achievable in the sense that if the sender Alice is able to wait long enough and send 
long strings of the input quantum states \(\rho_i\), then there exists a particular encoding and a decoding scheme that asymptotically attains 
the bound. Moreover, the encoding consists in collecting certain long and ``typical'' strings of the input states, and sending them all at once 
\cite{babarey, maarey}.

\subsection{Capacity of quantum dense coding}
\label{koto-capa-re?}

Suppose that Alice and Bob share a quantum state \(\rho_{AB}\). Alice performs the unitary operation \(U_i\) with probability \(p_i\), on her 
part of the state \(\rho_{AB}\). The classical information that she wants to send to Bob is \(i\). 
Subsequent to her unitary rotation, she sends her part of the state 
\(\rho^{AB}\) to Bob. Bob then has the ensemble \(\{p_i, \rho_i\}\), where 
\[\rho_i = U_i \otimes I \rho_{AB} U_i^\dagger \otimes I.\]

The information that Bob is able to gather is \(I_{acc}(\{p_i, \rho_i\})\). This quantity is bounded above by 
\(\chi(\{p_i, \rho_i\})\), and is asymptotically achievable. 
The ``one-capacity'' \(C^{(1)}\) of dense coding for the state \(\rho_{AB}\) is the Holevo bound for the best encoding by Alice:
\begin{equation}
\label{Kohinoor}
C^{(1)}(\rho) = \max_{p_i,U_i} \chi(\{p_i, \rho_i\}) \equiv \max_{p_i,U_i} \left(S(\overline{\rho}) - \sum_i p_i S(\rho_i)\right).
\end{equation}
The superscript \((1)\) reflects the fact that Alice is using the shared state once at a time, during the asymptotic process. 
She is not using entangled unitaries on more than one copy of her parts of the shared states \(\rho_{AB}\). As we will see below, 
encoding with entangled unitaries does not help her to send more information to Bob.

In performing the maximization in Eq. (\ref{Kohinoor}), first note that the second term in the right hand side (rhs) is 
\(-S(\rho)\), for all choices of the unitaries and probabilities. Secondly, we have 
\[S(\overline{\rho}) \leq S(\overline{\rho}_A) 
+ S(\overline{\rho}_B) \leq \log_2 d_A + S(\overline{\rho}_B),\] where 
\(d_A\) is the dimension of Alice's part of the Hilbert space of \(\rho_{AB}\), and 
 \( \overline{\rho}_A = \mbox{tr}_B \overline{\rho}\), \( \overline{\rho}_B = \mbox{tr}_A \overline{\rho}\). 
Moreover, \(S(\overline{\rho}_B) = S(\rho_B)\),
as nothing was done at Bob's end during the encoding procedure. (In any case, unitary operations does not change the spectrum, and hence 
the entropy, of a state.) Therefore, we have 
\[
\max_{p_i,U_i} S(\overline{\rho}) \leq \log_2 d_A + S(\rho_B).
\]
But the bound is reached by any 
complete set of orthogonal unitary operators
$\{W_j\}$, to be  chosen with equal probabilities, 
which 
satisfy the \emph{trace rule} $\frac{1}{d_A^2}\sum_j W_j^\dagger \Xi W_j=\mbox{tr}[\Xi]I$, 
for any operator
$\Xi$. Therefore, we have 
\[
C^{(1)}(\rho) = \log_2 d_A + S( \rho_B ) - S(\rho). 
\]

The optimization procedure above sketched essentially follows that in Ref. \cite{tin1}.
Several other lines of argument are possible for the maximization. One is given in Ref. \cite{dui1} (see also \cite{TajMahal}).  Another way to proceed is 
to guess where the maximum is reached (maybe from examples or by taking the most symmetric option), and then perturb the guessed result. If the 
first order perturbations vanish, the guessed result is correct, as the von Neumann entropy is a concave function and 
the maximization is carried out over a continuous parameter space. 

Without using the additional resource of entangled states, Alice will be able to reach a capacity of just 
\(\log_2d_A\) bits. Therefore, entanglement in a state \(\rho^{AB}\) is useful for dense coding if \(S( \rho^B ) - S(\rho)>0\). 
Such states will be called dense-codeable (DC) states. Such states exist, an example being the singlet state.

Note here that if Alice is able to use entangled unitaries on two copies of the shared state \(\rho\), 
the capacity is not enhanced (see Ref. \cite{ar-keu-korechhey?}). Therefore, the one-capacity is really the asymptotic
capacity, in this case. Note however that this additivity is known 
only in the case of encoding by unitary operations. A more general encoding may still have 
additivity problems (see e.g. \cite{Debu}). Here, 
we have considered unitary encoding only. This case is both mathematically more accessible, and experimentally more viable.


A bipartite state \(\rho^{AB}\) is useful for dense coding if and only if 
\(S( \rho^B ) - S(\rho)>0\). It can be shown that this relation cannot hold for PPT entangled states \cite{MarieCurie} (see also
\cite{TajMahal}).
Therefore a DC state is always NPT. 
However, the converse is not true: There exist states which are NPT, but they are not useful for dense coding. Examples of 
such states can be obtained by the considering the Werner state
\(\rho_p = p|\psi^-\rangle \langle \psi^- | + \frac{1-p}{4} I \otimes I\) \cite{Werner}.

\begin{figure}
\begin{center}
\includegraphics[width=8cm]{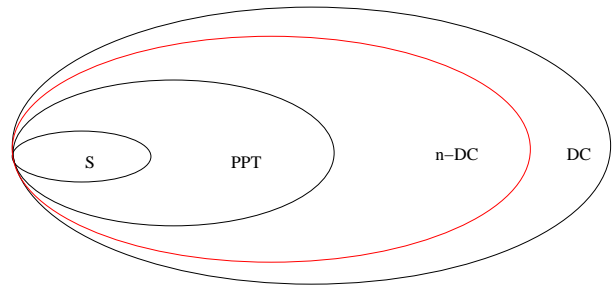}
\end{center}
\caption[Classification of bipartite quantum states according to their usefulness in dense coding]
{Classification of bipartite quantum states according to their usefulness in dense coding. The convex innermost 
region, marked as S, consists of separable states. The shell surrounding it, marked as PPT, is the set of 
PPT entangled states. The next shell, marked as n-DC, is the set of all states that are NPT, but not useful for dense 
coding. The outermost shell is that of dense-codeable states.}\label{fig:ghorar-dim}
\end{figure}

The discussions above leads to the following classification of bipartite quantum states: 
\begin{enumerate}
\item Separable states: These states are of course not useful for dense coding. They can be prepared by LOCC.

\item PPT entangled states: These states, despite being entangled, cannot be used for dense  coding. 
Moreover,
their entanglement cannot be detected by the  partial transposition criterion.

\item NPT non-DC states: These states are entangled, and their entanglement can be detected by the 
partial transposition criterion. However, they are not useful for dense coding. 

\item DC states: These entangled states can be used for dense coding. 

\end{enumerate}
The above classification is illustrated in figure \ref{fig:ghorar-dim}.
A generalisation of this classification has been considered in Refs. \cite{TajMahal,ar-keu-korechhey?}.

\section{Further reading: Multipartite states}

The discussion about detection of bipartite entanglement presented above is of course quite far from 
complete. For further reading, we have presented a small sample of references embedded in the text above. 
We prefer to conclude this chapter with a few remarks on multipartite states. 

The case of detection of entanglement of pure states is again simple. One quickly realizes that a multipartite pure state 
is entangled if and only if it is entangled in at least one bipartite splitting. 
So, for example, the state \(|\mbox{GHZ}\rangle = \frac{1}{\sqrt{2}} ( |000\rangle + |111\rangle ) \) \cite{GHZ} is 
entangled, because it is entangled in the A:BC bipartite splitting (as also in all others).

The case of mixed states is however quite formidable. In particular, the results obtained in the bipartite mixed state case, cannot 
be applied to the multiparty scenario. One way to see this is to notice the existence of states which are separable in any bipartite splitting,
while the entire state is entangled. An example of such a state is given in Ref. \cite{UPBPRL}. For further 
results about entanglement criteria, detection, and classification of multipartite states,  
see e.g. 
\cite{panch, saat, Mumbai, satattor, Horo-realign, sakkhiprl, ek, chhoi, char, tin, dui, sotero}, and 
references therein.

\section{Problems}

\noindent \textbf{Problem 1} Show that the singlet state has nonpositive partial transposition.

\noindent \textbf{Problem 2} Consider the Werner state  \(p |\psi^- \rangle \langle \psi^- | + (1 -p) I/4\) in \(2 \otimes 2\), where 
\(0\leq p \leq 1\) \cite{Werner}. Find the values of the mixing parameter \(p\),
for which entanglement in the Werner state can be detected by the partial transposition criterion. 

\noindent \textbf{Problem 3} Show that in \({\cal C}^2 \otimes {\cal C}^2\), the partial transposition of a 
density matrix can have at most one negative eigenvalue. 

\noindent \textbf{Problem 4} Given two random variables \(X\) and \(Y\), show that the Shannon entropy of the joint distribution cannot 
be smaller than that of either.

\noindent \textbf{Problem 5} Prove Theorem 5. 

\noindent \textbf{Problem 6} Prove Lemma 1. 

\noindent \textbf{Problem 7} Prove Theorem 10.

\noindent \textbf{Problem 8} Show that each of the shells depicted in figure \ref{fig:ghorar-dim} are nonempty, and of 
nonzero measure. Show also that all the boundaries are convex.

\printindex
\end{document}